\documentclass{amsproc}

\let\a=\alpha    \let\e=\epsilon
    
 \let\m=\mu \let\n=\nu

\let\C=\Chi

\def\nn{\nonumber} \def\bd{\begin{document}} \def\ed{\end{document}}
\def\ds{\documentstyle} \let\fr=\frac \let\bl=\bigl \let\br=\bigr
\let\Br=\Bigr \let\Bl=\Bigl
\let\bm=\bibitem
\let\na=\nabla
\let\pa=\partial \let\ov=\overline
\let\cal=\mathcal
\newcommand{\be}{\begin{equation}}
\newcommand{\ee}{\end{equation}}
\newcommand{\eeq}{\end{equation}}
\def\ba{\begin{array}}
\def\ea{\end{array}}
\def\ft#1#2{{\textstyle{{\scriptstyle #1}\over {\scriptstyle #2}}}}
\def\fft#1#2{{#1 \over #2}}
\def\del{\partial}
\def\vp{\varphi}
\def\st#1{{\scriptstyle #1}}
\def\sst#1{{\scriptscriptstyle #1}}
\def\bigsq{\mathord{\dalemb{9}{10}\hbox{\hskip1pt}}}
\def\oneone{\rlap 1\mkern4mu{\rm l}}
\def\td{\tilde}
\def\wtd{\widetilde}
\def\ie{\rm i.e.\ }
\def\dalemb#1#2{{\vbox{\hrule height .#2pt
         \hbox{\vrule width.#2pt height#1pt \kern#1pt
                 \vrule width.#2pt}
         \hrule height.#2pt}}}
\def\square{\mathord{\dalemb{6.8}{7}\hbox{\hskip1pt}}}
\def\cramp{\medmuskip = 2mu plus 1mu minus 2mu}
\def\cramper{\medmuskip = 2mu plus 1mu minus 2mu}
\def\crampest{\medmuskip = 1mu plus 1mu minus 1mu}
\def\uncramp{\medmuskip = 4mu plus 2mu minus 4mu}
\newcommand{\ho}[1]{$\, ^{#1}$}
\newcommand{\hoch}[1]{$\, ^{#1}$}
 \newcommand{\bea}{\begin{eqnarray}}
 \newcommand{\eea}{\end{eqnarray}}
\newcommand{\ra}{\rightarrow}
\newcommand{\lra}{\longrightarrow}
\newcommand{\Lra}{\Leftrightarrow}
\newcommand{\ap}{\alpha^\prime}
\newcommand{\bp}{\tilde \beta^\prime}
\newcommand{\tr}{{\rm tr} }
\newcommand{\Tr}{{\rm Tr} }
\def\0{{\sst{(0)}}}
\def\1{{\sst{(1)}}}
\def\2{{\sst{(2)}}}
\def\3{{\sst{(3)}}}
\def\4{{\sst{(4)}}}
\def\5{{\sst{(5)}}}
\def\6{{\sst{(6)}}}
\def\7{{\sst{(7)}}}
\def\8{{\sst{(8)}}}
\def\n{{\sst{(n)}}}
\def\cA{{{\cal A}}}
\def\cF{{{\cal F}}}
\def\tV{\widetilde V}
\def\tW{\widetilde W}
\def\tH{\widetilde H}
\def\tE{\widetilde E}
\def\tF{\widetilde F}
\def\tA{\widetilde A}
\def\im{{{\rm i}}}
\def\jm{{{\rm j}}}
\def\km{{{\rm k}}}
\def\tY{{{\wtd Y}}}
\def\ep{{\epsilon}}
\def\vep{{\varepsilon}}
\def\R{\rlap{\rm I}\mkern3mu{\rm R}}
\def\bD{{{\bar D}}}
\def\R{{{\Bbb R}}}
\def\C{{{\Bbb C}}}
\def\H{{{\Bbb H}}}
\def\CP{{{\Bbb C}{\Bbb P}}}
\def\RP{{{\Bbb R}{\Bbb P}}}
\def\Z{{{\Bbb Z}}}
\def\bA{{{\Bbb A}}}
\def\bB{{{\Bbb B}}}
\newcommand{\NP}{Nucl. Phys. }
 \newcommand{\upenn}{Department of Physics and Astronomy\\
 University of Pennsylvania, Philadelphia,  PA 19104, USA}
\newcommand{\itp}{Institute for Theoretical Physics\\
 University of California, Santa Barbara, CA 93106-4030}
\begin{document}

\begin{flushright}
\hfill{UPR--944--T ITP--01--49 MCTP--01--29 CTP~TAMU--23/01%
\footnote{Strings 2001 contribution, based on the talk given by 
 M. Cveti\v c}
}
\end{flushright}
\title{Resolved Branes and M-theory on Special Holonomy 
Spaces%
%\footnote{Talk presented by M. Cveti\v c.}
}
\author{Mirjam Cveti\v c, G.W. Gibbons, H. L\"u and C.N. Pope}
\address{\upenn \\ and \\ \itp}
%\address{Department of Physics and Astronomy,
%University of Pennsylvania,
%Philadelphia, PA 19104--6396, USA}
\email{cvetic@cvetic.hep.upenn.edu} 
\address{ DAMTP, Centre for Mathematical Sciences,
 Cambridge University\\ Wilberforce Road, Cambridge CB3 OWA, UK}
\email{G.W.Gibbons@damtp.cam.ac.uk}
\address{ Michigan Center for Theoretical Physics,
University of Michigan\\ Ann Arbor, MI 48109, USA}
\email{honglu@umich.edu}
\address{Center for Theoretical Physics,
Texas A\&M University, College Station, TX 77843, USA}
\email{pope@physics.tamu.edu}

\begin{abstract}
We review the construction of regular $p$-brane solutions of M-theory
and string theory with less than maximal supersymmetry whose
transverse spaces have metrics with special holonomy, and where
additional fluxes allow brane resolutions via transgression terms.  We
focus on the properties of resolved M2-branes and fractional D2-branes,
whose transverse spaces are Ricci flat eight-dimensional and
seven-dimensional spaces of special holonomy. We also review fractional
M2-branes with transverse spaces corresponding to the new two-parameter
$Spin(7)$ holonomy metrics, and their connection to fractional D2-branes
with transverse spaces of $G_2$ holonomy.
\end{abstract}

\maketitle

\section{Introduction}

Regular supergravity solutions with less than
maximal supersymmetry may provide viable gravity duals to
strongly coupled field theories with less than maximal
supersymmetry. In particular, the regularity of such solutions  at
small distances sheds light  on confinement and chiral symmetry
breaking in the infrared regime of  the dual strongly coupled
field theory
 \cite{klst}.

In this contribution we shall review the construction of such
regular supergravity solutions with emphasis on resolved
M2-branes of 11-dimensional supergravity  and fractional
D2-branes of Type IIA supergravity,  which provide viable
gravity duals of strongly coupled three-dimensional 
theories with ${\cal N}=2$ and ${\cal N}=1$ supersymmetry.

    This construction has been referred to as a ``resolution via
transgression'' \cite{clp1}.  It involves the replacement of the
standard flat transverse space by a smooth space of special holonomy,
i.e. a Ricci-flat space with fewer covariantly constant spinors.
Furthermore, additional field strength contributions are involved,
which are provided by harmonic forms in the space of special holonomy.
Transgression--Chern-Simons terms modify the equation of motion and/or
Bianchi identity for the original $p$-brane field strength. The
construction will be reviewed in Section 2 for the simpler prototype
example of a self-dual string in $D=6$ (0,1) supergravity, and then in
Section 3 the results for resolved M2-branes and D2-branes are
summarized.

   The explicit construction of such solutions has led to some new
mathematical developments, for example obtaining harmonic
forms for a large class of special holonomy metrics. As a prototype
example we shall review the construction of the metric and the
middle-dimensional forms for the Stenzel manifolds in $D=2n$ (with $n\ge
2$ integer) \cite{cglp1,cglp3} (Section 4a). We shall also disuss
examples of known $G_2$ holonomy spaces and their associated harmonic forms
(Section 4b).  In Section 4c we also discuss the old as well as the
new two-parameter metric with $Spin(7)$ holonomy
\cite{cglpspin7,cglpspin7m} and the associated harmonic forms (Section
4c).

  We then turn to the physics implications for resolved M2-branes
\cite{clp1,cglp3} (Section 5a) and fractional D2-branes
\cite{clp1,cglp2} (Section 5b).  In Section 5c we summarize the properties
of the fractional M2-brane whose transverse space is that of the new
$Spin(7)$ holonomy metrics, and the relationship to a fractional D2-brane
obtained as a reduction on the $S^1$ fiber of the $Spin(7)$ holonomy
metric \cite{cglpspin7}.

  In Section 6 we summarize the key results and spell out directions
for future work.

    The work presented in this paper was initiated in \cite{clp1}
and further pursued in a series of papers that provide both new
technical mathematical results and physics implications for resolved
$p$-brane configurations
\cite{cglp1,cglp3,cglpspin7,cglpspin7m,cglp2}.

\section{Resolution via Transgression}
\subsection{Motivation}
The $AdS_{D+1}/CFT_D$ correspondence \cite{ma,gkp,wi} provides a
quantitative insight into strongly coupled superconformal gauge
theories in $D$ dimensions, by studying the dual supergravity solutions.  The
prototype supergravity dual is the D3-brane of Type IIB theory, with the
classical solution
%%%%%
{ \bea ds_{10}^2 &=& H^{-1/2}\, dx\cdot dx +
H^{1/2}\, (dr^2 + r^2\,
d\Omega_5^2)\,,\nn\\
F_\5 &=& d^4x\wedge dH^{-1} + {\hat *(d^4x\wedge dH^{-1})}\,,\nn\\
\square H &=& 0\,\Rightarrow H = 1 + \fft{R^4}{r^4}\,\,\, .
 \label{D3N4}
\eea}
%%%%%%%
In the decoupling limit $H = 1 + \fft{R^4}{r^4}
\rightarrow  \  \fft{R^4}{r^4}$ this reduces to $AdS_5\times S^5$,
which provides a  gravitational dual  of  the strongly coupled
${\cal N}=4$ super-Yang-Mills  (SYM) theory.

   Of course, the ultimate goal of this program is to elucidate
strongly coupled YM theory, such as QCD, that has no
supersymmetry. But for the time being important steps have been
taken to obtain viable (regular) gravitational duals of strongly
coupled field theories with less than maximal supersymmetry. In
particular, within this framework we shall shed light on gravity duals
of field theories in $D=\{2,3,4\}$ with ${\cal N}=\{1,2\}$
supersymmetry.

 As a side comment, within $D=5$ ${\cal N}=2$ gauged supergravity
progress has been made (see \cite{becv,ceetal} and references therein)
in the explicit construction of domain wall solutions, both with
vector-multiplets {\it and} hyper-multiplets, which lead to smooth
solutions that provide viable gravity duals of $D=4$ ${\cal N}=1$
conformal field theories.  However, the aim in this contribution is to
discuss the higher dimensional embeddings and the interpretation of
such gravity duals.

   A typical way to obtain a supergravity solution with lesser
supersymmetry is to replace the flat transverse 6-dimensional space
$ds_6^2 =dr^2 + r^2\, d\Omega_5^2$ of the D3-brane in (\ref{D3N4})
with a smooth non-compact Ricci-flat space with fewer Killing spinors.
In this case the metric function $H$ still satisfies $ \square H =0$,
but now $\square$ is the Laplacian in the new Ricci-flat transverse
space.  This procedure ensures one has a solution with reduced
supersymmetry; however the solution for $H$ is singular at the inner
boundary of the transverse space, signifying the appearance of the
(distributed) D3-brane source there.

  A resolution of the singularity (and the removal of the additional
source) can take place if one turns on additional fluxes
(``fractional'' branes). Within the D3-brane context, the Chern-Simons
term of type IIB supergravity modifies the equations of motion:
%%%%%
\bea 
dF_\5& = &d{*F_\5} = F_\3^{\rm NS} \wedge F_\3^{\rm RR} =\ft1{2i}
F_\3 \wedge \bar F_\3\,,\nn\\ F_\3 &\equiv& F_\3^{\rm RR} + {\rm i}\,
F_\3^{\rm NS}=m L_\3\, ,
\label{D3cs}
\eea 
%%%%%
where $L_\3$ is a complex harmonic self-dual 3-form on the
6-dimensional Ricci-flat space. Depending on the properties of $L_3,$
this mechanism may allow for a smooth and thus viable supergravity
solution. This is precisely the mechanism employed by Klebanov and
Strassler, which in the case of the deformed conifold yields a
supergravity dual of $D=4$ ${\cal N}=1$ SYM theory.  (For related and
follow up work see, for example,
\cite{grpo,manu,gu,pats,bebe,bvflmp,ah,ca,gaetal}.  For earlier work
see, for example, \cite{klwi1,gukl,klne,klts}.)

   In a general context the resolution via transgression \cite{clp1} is a
consequence of the Chern-Simons-type (transgression) terms that are
ubiquitous in supergravity theories.  Such terms modify the Bianchi
identities and/or equations of motion when additional field strengths
are turned on.  $p$-brane configurations with $(n+1)$-transverse
dimensions, i.e.  with ``magnetic'' field strength $F_{(n)}$, can have
additional field strengths $F_{(p,q)}$ which, via transgression terms,
modify the equations for $F_{(n)}$:
%%%%%
\begin{equation}
dF_{(n)}= F_{(p)}\wedge F_{(q)}\, ; \ \ \  (p+q=n+1)\, .
\label{transgr} 
\end{equation}
%%%%%%
If the $(n+1)$-dimensional transverse Ricci-flat space admits a harmonic
$p$-form $L_{(p)}$ then the equations of motion are satisfied if one sets $
F_{(p)} =m L_{(p)}$, and by duality $F_{(q)}\sim \m *L_{(p)}\,
$.  Depending on the $L^2$ normalizability properties of $L_{(p)}$, one may
be able to obtain resolved (non-singular) solutions.  Let us now
illustrate the principle explicitly, first for a somewhat simpler
example than that of the D3-brane, namely the self-dual string.

\subsection{Self-dual string}
        The self-dual string in $D=6$, $(1,0)$ supergravity theory arises from
a Lagrangian of the form:
%%%%%
\begin{equation}
{\cal L} = \sqrt{g} (R - \ft1{12} F_\3^2)\, , \ \ F_\3 = {*F_\3}\,.
\end{equation}
The solution with a flat transverse space is
%%%%%%
\bea
ds_{6}^2 &=& H^{-1}\, (-dt^2 + dx^2) + H\, (dr^2 + r^2\, d\Omega_3^2)
\,,\nn\\
F_\3 &=& dt\wedge dx \wedge dH^{-1} + {\rm dual}\,,\nn\\
\square H &=& 0\,\, \Rightarrow \,\, H = 1 + \fft{R^2}{r^2}\,.\label{stringF}
\eea
%%%%%
In the decoupling limit we have $H = 1 + \fft{R^2}{r^2}\
\to \fft{R^2}{r^2}$, and the space-time becomes
$AdS_3\times S^3$.

A self-dual string with less supersymmetry is obtained by replacing
the four-dimensional flat transverse space with the
Eguchi-Hanson metric \cite{egha}:
%%%%%
\begin{equation}
 ds_4^2 = W^{-1}\, dr^2 +\ft14 r^2\, W\, (d\psi + \cos\theta\,
d\phi)^2
 + \ft14 r^2\, (d\theta^2 + \sin^2\theta\, d\phi^2)\,,
\label{EH}
\end{equation}
where
$W = 1-\fft{a^4}{r^4}$, and  $r=\{a,\infty\}$ and the Vierbeine take the form:
\bea
&&e^0 = W^{-1/2}\, dt\,,\qquad e^3 = \ft12 r\, W^{1/2}\,
(d\psi + \cos\theta\, d\phi)\,,\nn\\
&&e^1 = \ft12r\, d\theta\,,\qquad
e^2 = \ft12r\, \sin\theta\, d\phi\,.\label{EHV}
\eea
%%%%%
This metric is a resolution of the Ricci-flat cone over $S^3/\Z_2$. As
$r\to a$, the metric locally approaches  $\R^2\times S^2$, with $S^2$
being a minimal 2-surface or ``bolt.''

   The function $H$ is still harmonic, satisfying  
$\square H = 0$, but now the Laplacian is calculated in the
Eguchi-Hanson transverse space metric (\ref{EH}): 
\begin{equation} (r^3 W H')'
=0\,\,\Rightarrow H=1 +c_1\, \log(\fft{r^2+a^2}{r^2-a^2})\, . 
\end{equation}
%%%%%
As $r\to a$ the solution becomes singular, with the coefficient $c_1$
characterising the strength of a self-dual string source, distributed over
the $S^2$ bolt.

   The resolved self-dual string solution arises from the
modification of the Bianchi identity and equations of motion via
transgression terms (coming from a K3 reduction of the heterotic string.): 
%%%%%
\begin{equation} 
d F_\3 = d {*F_\3} = F_\2^i\wedge F_\2^i\,.
\end{equation}
%%%%%
Here
$F_2^{i}$  $(i=1,\cdots 16)$ are the gauge-field strengths in  the
Cartan subalgebra of $E_8\times E_8$ or $Spin(32)$.

The modified Ansatz with a single $U(1)$-gauge field turned on is
given by
%%%%%%%
\bea
ds_{6}^2 &=& H^{-1}\, (-dt^2 + dx^2) + H\, ds_4^2\,,\nn\\
F_\3 &=& dt\wedge dx \wedge dH^{-1} + {\rm dual}\,,\nn\\
F_\2 &=& m L_\2\,, \nn\\
\square H &=&-\ft12 L_\2^2\,,
\eea
%%%%
where $L_\2$ is a self-dual harmonic 2-form 
in the Eguchi-Hanson metric:
%%%%%
\begin{equation}
L_\2 = \ft{1}{r^4}\, (e^0 \wedge e^3 + e^1\wedge e^2)\,.
\end{equation}
%%%%%
Here the $e^i$ are a Vierbeine (\ref{EHV}) for Eguchi-Hanson
metric.  The harmonic 2-form $L_\2$ is $L^2$-normalizable!

  The modified solution is given by
%%%%%%
\begin{equation} 
H = 1 + \fft{m^2 + a^4\, b}{4a^6}\log(\fft{r^2-a^2}{r^2 +a^2})
+ \fft{m^2}{2a^4\, r^2}\,. 
\end{equation}
%%%%%%
By choosing the integration constant so that $b=-m^2/a^4$,
the singular source at the inner boundary of the transverse
space is eliminated,  yielding a {\it regular} solution: 
%%%%%
\begin{equation}
H=1+\fft{m^2}{2a^4\, r^2}\,. 
\end{equation}
%%%%% 
The solution is supersymmetric, and because $L_{(2)}$ is in $L^2$ it
is completely regular, with a well-defined ADM mass.

   A reduction on the $\psi$, $\theta$ and $\phi$
 coordinates of  the Eguchi-Hanson metric (\ref{EH})  yields in the
 decoupling limit ($H\rightarrow R^2/r^2$) a
 regular  domain wall solution in $D=3$:
%%%%%%
\begin{equation}
ds_3^2 = \fft{r^2}{R^2}\, W\, (-dt^2 + dx^2) + \fft{R^2\,
dr^2}{r^2}\,,
\end{equation}
%%%%%%
which asymptotically approaches $ AdS_3$ and provides a gravity dual
of 2-dimensional CFT theory with $1/2$ of maximal supersymmetry.  A
straightforward analysis in the IR (small $r$) regime reveals that the
bound-state spectrum for minimally coupled scalars is discrete, thus
indicating confinement.

\section{Summary of Resolved  M2-branes and D2-branes}

\subsection{M2-brane}
The transgression term in the 4-form field equation 
in 11-dimensional supergravity is given by
%%%%%
\be
d{*F_\4} = \ft12 F_\4\wedge F_\4\,,
\end{equation}
%%%%%
and the  modified  M2-brane Ansatz takes the form
%%%%%
\bea 
d\hat s_{11}^2 &=& H^{-2/3}\,
dx^\mu\, dx^\nu\, \eta_{\mu\nu} +
H^{1/3}\, ds_8^2\,,\nn\\
 F_\4 &=& d^3x\wedge dH^{-1} + m\, L_\4\,,
\label{m2sol}\eea
%%%%%
where $L_\4$ is a  harmonic self-dual 4-form  in the 8-dimensional
 Ricci-flat
 transverse space.  The  equation for $H$ is ten given by
%%%%%%%
\be 
\square H = -\ft1{48} m^2\, L_\4^2\,.\label{m2har} 
\end{equation}
%%%%%%

For related work see, for example, 
\cite{bebe,bebe0,duetal,hata,kbe,hekl}.

\subsection{D2-brane}
The transgression modification in the 4-form field equation in type IIA
supergravity is
\begin{equation}
d(e^{\ft12\phi}\, {\hat * F_4}) = F_\4\wedge F_\3\, ,
\end{equation}
and the modified D2-brane Ansatz takes the form:
%%%%%%
\bea
ds_{10}^2 &=& H^{-5/8}\, dx^\mu\, dx^\nu\, \eta_{\mu\nu} +
H^{3/8}\, ds_7^2\,,\nn\\
F_\4 &=& d^3x\wedge dH^{-1} + m\, L_\4\,,\qquad
F_\3 = m\, L_\3\,,\qquad \phi = \ft14\log H\,,\label{d2sol}
\eea
%%%%%%
where $G_\3$ is a harmonic 3-form in the Ricci-flat 7-metric
$ds_7^2$, and $L_\4={*L_\3}$, with $*$ the Hodge dual with respect to
the metric $ds_7^2$.  The function $H$ satisfies
%%%%%
\begin{equation}
 \square H = -\ft16 m^2 L_\3^2\,,\label{d2har} 
\end{equation}
%%%%%
where $\square$ denotes the scalar Laplacian with respect to the
transverse 7-metric $ds_7^2$.  Thus the deformed D2-brane solution is
completely determined by the choice of Ricci-flat 7-manifold, and the
harmonic 3-form  supported by it.

\subsection{ Other Examples}
In general the transgression terms  modify field equations or 
Bianchi identities as given in (\ref{transgr}), thus allowing
resolved branes with $(n+1)$ transverse dimensions for the
 following  additional examples in M-theory and string theory:
\begin{itemize}
\item (i) D0-brane: $d{*F_\2} = {*F_\4}\wedge F_\3$, 
\item (ii) D1-brane: $d{*F}_\3^{\rm RR} = F_\5\wedge F_\3^{\rm NS}$, 
\item (iii) D4-brane: $dF_\4 = F_\3\wedge F_\2$, 
\item (iv) IIA string: $d{*F_\3} = F_\4\wedge F_\4$, 
\item (v) IIB string: $d{*F}_\3^{\rm NS} = F_\5\wedge F_\3^{\rm RR}$, 
\item (vi) heterotic 5-brane:  $dF_\3 = F_\2^i\wedge F_\2^i$.
\end{itemize}

In what follows, we shall  focus on resolved M2-branes and
fractional D2-branes.  For details of other examples and their
properties, see e.g., \cite{clp1,cglp1,he}.

\section{Mathematical Developments}
The construction of resolved supergravity solutions necessarily
involves the explicit form of the metric on the Ricci-flat special
holonomy spaces. These  spaces  fall into the following classes:
\begin{itemize}
\item K\"ahler spaces in $D=2n$ dimensions ($n$-integer) with $SU(n)$
holonomy,  and two covariantly constant spinors. There are many
examples, with the Stenzel metric on $T^*S^n$ providing a prototype. They are
typically asymptotically conical (AC).
\item Hyper-K\"ahler spaces in $D=4n$ with  $Sp(n)$ holonomy, and
$n+1$ covariantly constant spinors. Subject to certain technical
assumptions, Calabi's metric on the co-tangent
bundle of $\CP^n$ is the only complete irreducible cohomogeneity one 
example \cite{dasw}.
\item  In $D=7$ there are exceptional  $G_2$ holonomy  spaces with  one
covariantly constant spinor. Until recently only three AC examples
were known \cite{brysal,gibpagpop}, but
 new metrics have been recently constructed in \cite{bggg,cglp5}.
\item  In $D=8$ there are exceptional $Spin(7)$ holonomy spaces with  one
covariantly constant spinor; until recently only one AC example
was known \cite{brysal,gibpagpop}. New  metrics were recently
constructed in \cite{cglpspin7,cglpspin7m}.
\end{itemize}

  The focus is on a construction of  cohomogeneity one spaces that
are typically asymptotic to cones over Einstein spaces. Recent mathematical
developments evolved along two directions: (i) construction of
harmonic forms on known Ricci-flat spaces (see in particular
\cite{cglp1,cglp2}), (ii) construction of  new exceptional
holonomy spaces \cite{cglpspin7,cglpspin7m,cglp5,bggg}. In the
following  two subsections we  illustrate these developments by
 summarizing (i)   results on  the construction of
harmonic forms on  the  Stenzel   metric  \cite{cglp1} and $G_2$
holonomy metrics \cite{clp1,cglp2}, and (ii)
results for the  construction of new $Spin(7)$ two-parameter
metrics \cite{cglpspin7,cglpspin7m}, all serving as prototype examples.

\subsection{Harmonic forms for the Stenzel metric}
The Stenzel\cite{st} construction provides a
class of complete non-compact
Ricci-flat K\"ahler manifolds,   one for each even dimension,
on the co-tangent bundle of the $(n+1)$-sphere, $T^\star S^{n+1}$.
These are asymptotically conical, with principal orbits that
are described by the coset space $SO(n+2)/SO(n)$, and they have real
dimension $d=2n+2$.

\subsubsection{Stenzel metric}

In the following we summarize the relevant results for the
construction of the Stenzel metric. (For more details see
\cite{cglp1}.)  This construction \cite{cglp1,st} of the Stenzel
metric starts with $L_{AB}$, which are left-invariant 1-forms on the
group manifold $SO(n+2)$. By splitting the index as $A=(1,2,i)$, we
have that $L_{ij}$ are the left-invariant 1-forms for the $SO(n)$
subgroup, and so the 1-forms in the coset $SO(n+2)/SO(n)$ will be
%%%%%
\be
\sigma_i \equiv L_{1i}\,,\qquad \td\sigma_i \equiv L_{2i}\,,\qquad
\nu \equiv L_{12}\,.
\end{equation}
%%%%%
  The metric Ansatz takes the form:
%%%%%
\begin{equation} 
ds^2 = dt^2 + a^2 \sigma_i^2 + b^2\, \td\sigma_i^2 + c^2\,
\nu^2\,, \label{stenmeth}\end{equation}
%%%%%
where $a$, $b$ and $c$ are functions of the radial coordinate $t$. One defines
Vielbeine
%%%%%
\begin{equation}
e^0=dt\,,\qquad e^i = a\, \sigma_i\,,\qquad e^{\td i} = b\,
\td\sigma_i\,,\qquad e^{\td 0}=c\, \nu\, , \label{vielbein}
\end{equation} 
%%%%%
for which one can introduce a holomorphic
tangent-space basis of complex 1-forms $\ep^\a$:
%%%%%
\begin{equation}
\ep^0 \equiv -e^0 + \im\, e^{\td 0}\,,\qquad \ep^i = e^i + \im\,
e^{\td i}\,.\label{complexbasis}
\end{equation} 

Defining $a=e^\a$, $b=e^\beta$, $c=e^\gamma$, and introducing the new
coordinate $\eta$ by $a^n\, b^n\, c\, d\eta = dt$, one finds
\cite{cglp1} that the Ricci-flat equations can be obtained from a
Lagrangian $L=T-V$ which can be written as a ``supersymmetric
Lagrangian'': $L=\ft12 g_{ij}\, (d{\a^i}/d\eta)\, (d{\a^j}/d\eta) -V$,
where the potential can be written in terms of a superpotential, as
%%%%%
\be
V = - \ft12 g^{ij} \, \fft{\del W}{\del \a^i}\, \fft{\del W}{\del \a^j}
\end{equation} 
%%%%%
with
%%%%%
\be
W = \ft12 (a\, b)^{n-1}\, (a^2 + b^2 + c^2)\,.
\end{equation} 
%%%%%
Here $\a^i=(\a,\beta,\gamma)$. The
second-order equations for Ricci-flatness are satisfied if the
first-order equations $d{\a^i}/d\eta = \mp g^{ij}\, \del_j W$ are satisfied.
Thus we arrive at the first-order equations
%%%%%
\be
\dot a = \ft12 b^{-1}c^{-1} (b^2+c^2-a^2)\, , \ \
\dot b = \ft12 a^{-1}c^{-1} (a^2+c^2-b^2)\, , \ \
\dot c = \ft12 a^{-1}b^{-1} (a^2+b^2-c^2)\, , \ \
\label{firstorder}
\end{equation} 
where the dot again denotes the radial derivative $d/dt$
%%%%%

The explicit solution of (\ref{firstorder}) can be written in the form:
%%%%%
\bea
&&a^2\equiv e^{2\a} = R^{1/(n+1)}\, \coth r\,,\nn\\
&&b^2\equiv e^{2\beta} = R^{1/(n+1)}\, \tanh r\,,\\
&&
h^2=c^2\equiv e^{2\gamma} = \fft1{n+1}\, R^{-n/(n+1)}\, (\sinh
2r)^n\,,\nn
\eea
%%%%%
where
%%%%%
\be
R(r) \equiv  \int_0^r (\sinh 2u)^n\, du\,.\label{rdef}
\end{equation}
%%%%% 
Here the radial coordinate  $r$ is introduced as $dt=h\, dr$.

  For each $n$ the result is expressible in relatively simple terms.
For example, one finds
%%%%%
\be
n=1: R = \sinh^2 r\,;\ \
n=2: R =  \ft18(\sinh 4 r- 4r)\,;\ \
n=3: R = \ft23 (2+\cosh 2r)\, \sinh^4 r\, .
\end{equation} 
%%%%%
The case $n=1$ is the Eguchi-Hanson metric \cite{egha},
and for $n=2$ it is the deformed conifold \cite{dloc}.

 As $r$ approaches zero, the metric takes the form
%%%%%
\be
ds^2\sim dr^2+r^2{\tilde \sigma}_i^2 +\sigma_i^2+\nu^2\,,
\end{equation}
%%%%% 
which has the structure locally of the product  
$\R^{n+1}\times S^{n+1}$,  with $S^{n+1}$ being a  ``bolt.''
As $r$ tends to infinity, the metric becomes
%%%%%
\be
ds^2\sim d\rho^2 +\rho^2\{\textstyle{{n^2\over {(n+1)^2}}}\nu^2+
\textstyle{{n\over {2(n+1)^2}}}(\sigma_i^2 +{\tilde \sigma_i}^2)\}\, ,
\end{equation}
%%%%% 
representing a cone over the Einstein space $SO(n+2)/SO(n)$.

\subsubsection{Harmonic middle-dimension $(p,q)$ forms}

An Ansatz compatible with the symmetries of the Stenzel metric is of the form:
%%%%%%
\bea
L_{\sst{(p,q)}} &=& f_1\,\ep_{i_1\cdots i_{q-1}j_1\cdots j_{p}}\,
\bar \ep^0\wedge \bar \ep^{i_1}\wedge\cdots \wedge\bar\ep^{i_{q-1}}\wedge
\ep^{j_1}\wedge\cdots\wedge\ep^{j_{p}}\nn\\
&&+f_2\,\ep_{i_1\cdots i_{p-1}j_1\cdots j_{q}}\,
\ep^0\wedge \ep^{i_1}\wedge\cdots\wedge \ep^{i_{p-1}}\wedge
\bar \ep^{j_1}\wedge\cdots\wedge \bar \ep^{j_q}\,,
\label{middled}
\eea
%%%%%
with $f_1$, $f_2$ being functions of $r$, only.
%%%%%
The harmonicity condition becomes
$ dL_{\sst{(p,q)}}=0\, , $
since $ {*L_{\sst{(p,q)}}} = \im^{\, p-q}\,
L_{\sst{(p,q)}}\,$.  The functions
 $f_1$, $f_2$ are solutions of coupled first-order
 homogeneous differential equations, yielding a solution
that is finite as $r\to 0$:
%%%%%
\be
f_1 =\, q\, _2F_1\left[ \ft12 p, \ft12 (q+1), \ft12 (p+q) +1;
-(\sinh 2r)^2\right]\, ,
\end{equation} 
\be
f_2 = -\, p\, _2F_1\left[ \ft12 q, \ft12 (p+1), \ft12 (p+q) +1;
-(\sinh 2r)^2\right]\, .
\end{equation}
%%%%% 
For any
specific integers $(p,q)$, these are elementary functions of $r$.

For the two special cases of greatest interest, they have the following
  properties:

\begin{itemize}
\item $(p,p)$-forms in $4p$-dimensions: $f_1=-f_2={p\over
(\cosh{r})^{2p}}$ with
$|L_{(p,p)}|^2={const.\over {(\cosh{r})^{4p}}}$
 falls-off fast enough as $r\to\infty$.  This turns out to be the
 only $L^2$ normalizable form!

\item $(p+1,p)$-forms in $(4p+2)$-dimensions. As  $r\to \infty$:
$|L_{(p+1,p)}|^2\sim {1\over {[\sinh{(2r)}]^{2p}}}$ which is  marginally
$L^2$-non-normalizable.
\end{itemize}

   As for physics implications, the case in
$2(n+1)=4$ dimensions with an $L^2$-normalizable $L_{(1,1)}$-form
is  precisely the example of the resolved self-dual string
discussed in detail  in Section 2.2.

 In $2(n+1)=6$ dimensions, the $L_{(2,1)}$-form was constructed in
 \cite{klst}, and provides a resolution of the D3-brane. Since
 $L_{(2,1)}$ is only {\it marginally non-normalizable} as $r\to\infty$,
the decoupling limit of the space-time does not give an AdS$_5$, but
instead there is a logarithmic modification. In particular, this
modification accounts for a renormalization group running of the
difference of the inverse-squares of the two gauge group couplings in
the dual $SU(N)\times SU(N+M)$ SYM \cite{klne}.

 On the other hand in $2(n+1)=8$ dimensions the $L^2$ normalizable
$L_{(2,2)}$-form supports additional fluxes that resolve the original
$M2$-brane. The details of the explicit solution, as well as general
properties of M2-branes, will be given in Section 5.1.

   It turns out that one can construct regular resolved M2-branes for
many other examples of 8-dimensional special holonomy
transverse spaces. In particular, examples  with the original
$Spin(7)$ holonomy  transverse space \cite{clp1}, a number of new
K\" ahler spaces \cite{clp1,cglp2}, and  hyper-K\"ahler spaces
\cite{cglp3}, can all support $L^2$ normalizable 4-forms and
therefore can give rise to resolved supersymmetric M2-brane  solutions.

\subsection{Old $G_2$ holonomy metrics and their harmonic forms}
\subsubsection{ Resolved cones over $S^2\times S^4$  and $S^2\times \CP^2$}
The
first type of complete Ricci-flat 7-dimensional metrics of $G_2$
holonomy, obtained in \cite{brysal,gibpagpop}, correspond to
$R^3$ bundles over four-dimensional quaternionic-K\"ahler
Einstein base manifolds $M$.  These spaces are  of cohomogeneity
one, with level surfaces that are $S^2$ bundles over $M$ (also
known as ``twistor spaces'' over $M$). There are two cases that arise,
with $M$ being $S^4$ or $\CP^2$.  Thus the two manifolds have
level surfaces that are $\CP^3$ ($S^2$ bundle over $S^4$) or the
flag manifold $SU(3)/(U(1)\times U(1))$ ($S^2$ bundle over
$\CP^2$), respectively. These two manifolds are the bundles of
self-dual 2-forms over $S^4$ or $\CP^2$ respectively. They
approach $\R^3\times S^4$ or $\R^3\times \CP^2$ locally near the
origin. (The calculations for the two cases, with the principal
orbits being $S^2$ bundles either over $S^4$ or over $\CP^2$,
proceed essentially identically.)

%The metric is of the form:
 The complete Ricci-flat 7-metric on the bundle of self-dual
2-forms over $S^4$ was constructed in \cite{brysal,gibpagpop}. In
the notation of \cite{gibpagpop}, it is given by
%%%%%
\be d\hat s_7^2 = h^2\, dr^2 + a^2\, (D\mu^i)^2 + b^2\,
d\Omega_4^2\,, \end{equation} 
%%%%%
where $\mu^i$ are coordinates on $\R^3$ subject to $\mu^i\,
\mu^i=1$; $d\Omega_4^2$ is the metric on the unit 4-sphere, with
$SU(2)$ Yang-Mills instanton potentials $A^i$, and
%%%%%
\be D\mu^i \equiv d\, \mu^i + \ep_{ijk}\, A^j\, \mu^k\,. \end{equation} 
%%%%%
The field strengths $J^i\equiv dA^i + \ft12 \ep_{ijk}\, A^j\wedge
A^k$ satisfy the algebra of the unit quaternions,
$J^i_{\a\gamma}\, J^j_{\gamma\beta} = -\delta_{ij}\,
\delta_{\a\beta} + \ep_{ijk}\, J^k_{\a\beta}$.  A convenient
orthonormal basis is
%%%%%
\be \hat e^0 = h\, dr\,,\qquad \hat e^i = a\, D\mu^i\,,\qquad
\hat e^\a = b\, e^\a\,. \end{equation} 
%%%%%
(Note that although $i$ runs over 3 values, there are really only
two independent Vielbeine on the 2-sphere, because of the
constraint $\mu^i\, \mu^i=1$.)

    Constructing  fractional D2-branes requires finding a suitable
harmonic 3-form, which is square-integrable at short distance and
whose dual 4-form has a non-vanishing flux integral at infinity.
In fact a fully $L^2$-normalizable harmonic 3-form in this
example exit was obtained  in \cite{cglp3} and is given by
%%%%%
\bea L_\3 = &\ft12 u_1\, \mu^i\, \ep_{ijk} \, \hat e^0\wedge \hat
e^j \wedge \hat e^k \nn\\
+ &\ft12 u_2\, \mu^i\, J^i_{\a\beta}\, \hat
e^0\wedge \hat e^\a\wedge \hat\e^\beta + \ft12 J^i_{\a\beta}\,
u_3\, \hat e^i\wedge\hat e^\a\wedge \hat e^\beta\,,\label{vform}
\eea 
%%%%%
where the harmonicity conditions $dG_\3=0$ and $d{*G_\3}=0$
impose first-order equations on the functions $u_{1,2,3}$.  There
is a regular solution, given by
%%%%%
\bea
u_1 &=& \fft1{r^4} + \fft{P(r)}{r^5\, (r^4-1)^{1/2}}\,,\nn\\
u_2 &=& -\fft1{2(r^4-1)} + \fft{P(r)}{r\, (r^4-1)^{3/2}}\,,\label{usol}\\
u_3 &=& \fft1{4 r^4\, (r^4-1)} - \fft{(3 r^4 -1)\, P(r)}{
        4 r^5\, (r^4-1)^{3/2}}\,,\nn
\eea
%%%%%
where
%%%%%
\be P(r) = F(\ft12 \pi|-1) - F(\arcsin(r^{-1})|-1) =
\int_{\arcsin(r^{-1})}^{\fft12\pi} \fft{d\theta}{\sqrt{ 1+
\sin^2\theta}} \,. \end{equation} 
%%%%%
Note that the functions $u_i$ satisfy $u_1+2 u_2 + 4 u_3=0$,
which turns out to be a key condition required for
compatibility with the supersymmetry constraints for fractional
D2-branes. The asymptotic  formm  of (\ref{vform}) reveals that
$|L_{(3)}|^2=6(u_1^2+2u_2^2+6u_3^2)$ tends to a constant at small
$r$, and $|L_\3|^2\sim 1/r^8$ at large $r$. Since the metric has
$\sqrt{\hat g} \sim r^6$ at large $r$, it follows that the
harmonic 3-form {\it is $L^2$-normalizable}.

\subsubsection {Resolved cone over $S^3\times S^3$} The remaining
complete 7-dimensional manifold of $G_2$
holonomy obtained in \cite{brysal,gibpagpop} is again
of cohomogeneity one, with principal orbits are
topologically $S^3\times S^3$.  The manifold is the spin bundle
of $S^3$; near the origin it approaches locally $\R^4\times S^3$.
(Topologically, it is in fact just $\R^4\times S^3$, since the bundle
is topologically trrivial.)
%The metric is of the form:

  The Ricci-flat metric on the spin bundle of $S^3$ is given by
\cite{brysal,gibpagpop}
%%%%%
\be ds_7^2 = \a^2\, dr^2 + \beta^2\, (\sigma_i - \ft12
\Sigma_i)^2 + \gamma^2\, \Sigma_i^2\,,\label{met2} \end{equation} 
%%%%%
where the functions $\a$, $\beta$ and $\gamma$ are given by
%%%%%
\be \a^2 = \Big(1-\fft{1}{r^3}\Big)^{-1}\,,\qquad \beta^2 = \ft19
r^2\,  \Big(1-\fft{1}{r^3}\Big)\,,\qquad \gamma^2 = \ft1{12}
r^2\,. \end{equation} 
%%%%%
Here $\Sigma_i$ and $\sigma_i$ are two sets of left-invariant
1-forms on two independent $SU(2)$ group manifolds.  The level
surfaces $r=$constant are therefore $S^3$ bundles over $S^3$.
This bundle is trivial, and so in fact the level surfaces are
topologically $S^3\times S^3$.  The radial coordinate runs from
$r=a$ to $r=\infty$.  We define an orthonormal frame by
%%%%%
\be e^0= \a\, dr\,,\qquad e^i =\gamma\, \Sigma_i\,,\qquad e^{\td
i} = \beta\, \nu_i \end{equation} 
%%%%%
where $\nu_i\equiv \sigma_i -\ft12 \Sigma_i$.

 The metric (\ref{met2}) admits the following regular 
harmonic 3-form \cite{clp1}:
%%%%%
\be L_{(3)}= v_1\, e^0\wedge e^{\td i}\wedge e^i + v_2 \,
\ep_{ijk}\, e^{\td i} \wedge e^{\td j}\wedge e^k + \ft16 v_3\,
\ep_{ijk}\, e^i\wedge e^j \wedge e^k\,,\label{sec3form} \end{equation} 
%%%%%
where the functions $v_1$, $v_2$ and $v_3$ are given by
%%%%%
\be v_1= -\fft{ (3r^2 + 2r+1)}{r^4\, (r^2+r+1)^2}\,,\quad
 v_2= \fft{ (r^2 + 2r+3)}{r\, (r^2+r+1)^2}\,,\quad
v_3= \fft{ 3(r+1)(r^2+1)}{r^4\, (r^2+r+1)}\,.\label{vsol} \end{equation} 
%%%%%
The functions $v_{(1,2,3)}$  satisfy the relation $3v_1 -3v_2 +
v_3=0\,\label{vrel}$, which is crucial for establishing
that $L_{(3)}$ is compatible with supersymmetry of the associated
deformed D2-brane.
The  harmonic 3-form is square-integrable at short distance, but
gives a linearly divergent integral at large distance.
As shown in \cite{clp1}, the short-distance square-integrability
is enough to give a perfectly regular deformed D2-brane solution,
even though $L_{(3)}$ is {\it not $L^2$-normalizable}.

 \subsection{New $Spin(7)$ holonomy metrics}
\subsubsection{The old metric and harmonic 4-forms}

Until recently only one explicit example of a complete non-compact
metric on a $Spin(7)$  holonomy space was known
\cite{brysal,gibpagpop}. The principal orbits are $S^7$,
viewed as an $S^3$ bundle over $S^4$. The solution
(\ref{spin7metric}) is asymptotic to a cone over the ``squashed''
Einstein 7-sphere, and it approaches $\R^4\times S^4$ locally at
short distance (\ie $r\approx\ell$). The metric is of the form:
%%%%%
\be 
ds_8^2 = \Big(1- \fft{\ell^{10/3}}{r^{10/3}}\Big)^{-1} \, dr^2
         + \ft{9}{100}\, r^2\, \Big(1- \fft{\ell^{10/3}}{r^{10/3}}\Big)\,
         h_i^2 + \ft{9}{20} r^2\, d\Omega_4^2\,,\label{spin7metric}
\end{equation} 
%%%%%
where
$h_i\equiv \sigma_i - A_\1^i\,,$ and the
 $\sigma_i$ are left-invariant 1-forms on $SU(2)$, $d\Omega_4^2$ is
the metric on the unit 4-sphere, and $A_\1^i$ is the $SU(2)$
Yang-Mills instanton on $S^4$.  The $\sigma_i$ can be written in terms
of Euler angles as
%%%%%
\crampest
\be
\sigma_1 = \cos\psi\, d\theta + \sin\psi\, \sin\theta\,
d\varphi\,,\quad
\sigma_2 = -\sin\psi\, d\theta + \cos\psi\, \sin\theta\,
d\varphi\,,\quad
\sigma_3 = d\psi + \cos\theta\, d\varphi\,.\label{1forms}
\end{equation} 
\uncramp
%%%%%
A regular harmonic 4-form in this metric was obtained in \cite{clp1}
 and  is given in terms of a 3-form potential $B_{(3)}$ by
 $L_{(4)}=d B_{(3)}$. The 3-form $B_{(3)}$ is given by
%%%%%
 \be B_{(3)}=f\,h_1\wedge h_2\wedge h_3 +g\,h_i\wedge F^i
 \end{equation} 
 where
 \be
 f=\textstyle{1\over 5}\left({\ell\over r}\right)^{2/3}\left[\left({\ell\over
 r}\right)^{10/3}-1\right]\, , \ \ \ g=\left({\ell\over
 r}\right)^{2/3}\, ,
 \end{equation}
%%%%% 
yielding an $L^2$-normalizable 4-form.

\subsubsection{New $Spin(7)$ holonomy metric}

 The generalization that we shall consider involves allowing the
$S^3$ fibers of the previous construction themselves to be
``squashed.'' Namely, the $S^3$ bundle is itself written as a
$U(1)$ bundle over $S^2$ leading to the following ``twice
squashed'' Ansatz:
%%%%%
\be 
d\hat s_8^2 = dt^2 + a^2\, (D\mu^i)^2 + b^2\, \sigma^2 + c^2\,
d\Omega_4^2\,,\label{8ans} 
\end{equation}
%%%%% 
where $a$, $b$ and $c$ are
functions of the radial variable $t$. (The previous $Spin(7)$
example has $a=b$.) Here 
%%%%%
\be 
\mu_1 = \sin\theta\,
\sin\psi\,,\qquad \mu_2= \sin\theta\, \cos\psi\,,\qquad \mu_3=
\cos\theta\,, 
\end{equation}
%%%%%
 are the $S^2$ coordinates, subject to the
constraint $\mu_i\mu_i=1$, and
%%%%%
\be 
D\mu^i\equiv d\mu^i
+\ep_{ijk}\, A_\1^j\, \mu^k\,,\quad \sigma \equiv d\varphi +
\cA_\1\,,\quad \cA_\1 \equiv \cos\theta\, d\psi - \mu^i\,
A_\1^i\, , \label{kkvector} 
\end{equation} 
%%%%%
where the  field strength ${\cal
F}_{(2)}$ of the
 $U(1)$ potential ${\cal A}_{(1)}$ turns out to be given by:
$
\cF_\2 = \ft1{2} \ep_{ijk}\, \mu^k\, D\mu^i \wedge
D\mu^j - \mu^i\, F_\2^{i}$.

   The Ricci-flatness conditions can be satisfied by solving the first-order
equations of a supersymmetric Lagrangian, yielding the
following special solution (for details see
\cite{cglpspin7,cglpspin7m}):
%%%%% 
\bea 
ds_8^2 = &\fft{(r-\ell)^2\,
dr^2}{(r-3\ell)(r+\ell)} + \fft{\ell^2\,
(r-3\ell)(r+\ell)}{(r-\ell)^2}\, \sigma^2 +\nn\\
&\ft14(r-3\ell)(r+\ell)\, (D\mu^i)^2 + \ft12(r^2-\ell^2)\,
d\Omega_4^2\,, \label{sol2} 
\eea
%%%%%
 The quantity $\ft14[\sigma^2 + (D\mu^i)^2]$ is the metric on the unit
3-sphere, and so in this case we find that the metric smoothly
approaches $\R^4\times S^4$ locally, at small distance ($r\to
3\ell$), i.e. it has the same topology as the old $Spin(7)$
holonomy space. On the other hand, it locally 
approaches ${\cal M}_7\times S^1$ at large distance.   Here
${\cal M}_7$ denotes the 7-manifold of $G_2$ holonomy that is the
$\R^3$ bundle over $S^4$ \cite{brysal,gibpagpop}. Asymptotically
the  new metric behaves like a circle bundle over an
asymptotically conical manifold in which the length of the $U(1)$
fibers tends to a constant; in other words, it is ALC.

If one takes $r$ to be negative, or instead analytically continues the
solution so that $\ell \to -\ell$ (keeping $r$ positive), one gets a
different complete manifold. Thus instead of (\ref{sol2}), the
quantity $\ft14(\sigma^2+(D\mu^i)^2 + d\Omega_4^2)$ is precisely the
metric on the unit 7-sphere, and so as $r$ approaches $\ell$ the
metric $ds_8^2$ smoothly approaches $\R^8$.  At large $r$ the
function $b$, which is the radius in the $U(1)$ direction $\sigma$,
approaches a constant, and so the metric tends to an $S^1$ bundle
over a 7-metric of the form of a cone over ${\Bbb C} {\Bbb P}^3$ ; it
has the same asymptotic form as (\ref{sol2}).  The manifold in this
case is topologically $\R^8$.

   In \cite{cglpspin7,cglpspin7m} the general solution to the
first-order system of equations is obtained, leading to additional
families of regular metrics of $Spin(7)$ holonomy, which are complete
on manifolds $\bB_8^\pm$ that are similar to $\bB_8$. These additional
metrics have a non-trivial integration constant which parameterizes
inequivalent solutions. (For details see \cite{cglpspin7m} and
Appendix A of \cite{cglpspin7}).

\subsubsection{$L^2$-normalizable harmonic 4-forms in the 
new $Spin(7)$ metrics}

     $L^2$ normalizable harmonic 4-forms
for the new $Spin(7)$ 8-manifolds were obtained in \cite{cglpspin7}.
The starting point is the following Ansatz:
%%%%%%
\bea 
L_{(4)} &=& u_1\, (h\, a^2\, b\, dr\wedge \sigma\wedge X_\2
\pm c^4\, \Omega_\4) + u_2\, (h\, b\, c^2\, dr\wedge \sigma\wedge
Y_\2 \pm a^2\,
c^2\, X_\2\wedge Y_\2)\nn\\
&&+ u_3\, (h\, a\, c^2\, dr\wedge Y_\3 \mp b\, a\,
c^2\,\sigma\wedge X_\3)\,,\label{4formans} 
\eea
%%%%%%
where $\Omega_\4$ is the volume form of the unit $S^4$, and
%%%%%
\bea 
&&X_\2\equiv \ft12\epsilon_{ijk}\, \mu^i\, D\mu^j\wedge
D\mu^k\,,\qquad
X_\3\equiv D\mu^i\wedge F_\2^i\,,\nn\\
&&Y_\2\equiv\mu^i\,F_\2^i\,,\qquad Y_\3\equiv\epsilon_{ijk}\,
\mu^i\, D\mu^j\wedge F_\2^k\,.\label{23forms} 
\eea
%%%%%
The upper and lower sign choices in (\ref{4formans}) correspond to
self-dual and anti-self-dual 4-forms respectively.  Here  a radial
coordinate $r$ is introduced, which is related to $t$ by $dt=h\, dr$.

  The harmonicity $dL_{(4)}=0$ of $L_{(4)}$ implies first-order
coupled differential equations for the functions $u_1$, $u_2$ and
$u_3$ (for details see \cite{cglpspin7}).  For the metric
(\ref{sol2}), the regular solution is given by
%%%%%
\bea 
&&u_1=\fft{2(r^4+8r^3+34r^2-48r+21)}{(r-1)^3(r+1)^5}\,\qquad
u_2= - \fft{r^4+4r^3-18r^2+52r-23}{(r-1)^3(r+1)^5}\,,\nn\\
&&u_3=\fft{2(r^2+14r-11)}{(r-1)^2(r+1)^5}\,.\label{basd} 
\eea
%%%%%
(For the sake of simplicity we set $\ell =1$.)  It can be seen
that $L_{(4)}$, whose norm is
$|L_{(4)}|^2=48(u_1^2+2u_2^2+4u_3^2)$, is in $L^2$.

For the manifold corresponding to the analytic continuation to
$\ell\to -\ell$ the resulting  anti-selfdual 4-form is specified
by the following functions:
%%%%%
\be 
u_1 =\fft{2}{(r+1)^3(r+3)}\,,\quad
u_2 =-\fft{r^2 +10r + 13}{(r+1)^3(r+3)^3}\,,\quad u_3
=-\fft2{(r+1)^2(r+3)^3}\,.\label{aasd} 
\end{equation}
%%%%%%
(Again for the sake of simplicity we set $\ell=1$.) The
resulting 4-form is again $L^2$-normalizable.

     Both of the above harmonic anti-self-dual 4-forms (\ref{aasd})
and (\ref{basd}) satisfy the linear relation
%%%%%
$ u_1+2u_2-4u_3=0\,,$
%%%%
which ensures the supersymmetry of the resolved brane
configurations.  
(Explicit expressions for the self-dual forms have also been obtained in
\cite{cglpspin7}, but these do not yield supersymmetric 
resolved brane configurations.)

\section{Applications}
\subsection{Resolved M2-branes}
\subsubsection{Resolved M2-brane with  transverse Stenzel space}
In this subsection, we shall summarize the  construction of  a
deformed M2-brane using the 8-dimensional Stenzel metric for the
transverse $ds_8^2$.  In this case, the index $i$ on $\sigma_i$
and $\td\sigma_i$ in the metric (\ref{stenmeth}) runs over 3
values. The Ricci-flat solution coming from the first-order
equations (\ref{firstorder}) is given by
%%%%%
\bea 
&&a^2= \ft13 (2+\cosh 2r)^{1/4}\, \cosh r\,,\qquad
b^2 =  \ft13 (2+\cosh 2r)^{1/4}\, \sinh r\, \tanh r\,,\nn\\
&&h^2=c^2 =  (2+\cosh 2r)^{-3/4}\, \cosh^3 r\,,\label{d8met} 
\eea
%%%%%
with the metric then given by (\ref{stenmeth}).  The radial
coordinate runs from $r=0$ to $r=\infty$, and the metric lives on
a smooth complete non-compact manifold.

 The $L^2$-normalizable self-dual harmonic 4-form (of type $(2,2)$)is
 a special case of (\ref{middled}) and is given by:
%%%%%
\bea L_{(2,2)} &=& \fft3{\cosh^4 r}\, [e^{\td 0 }\wedge e^1\wedge
e^2\wedge e^3
+ e^0\wedge e^{\td 1}\wedge e^{\td 2} \wedge e^{\td 3}] \nn\\
&&+ \fft1{2\cosh^4 r}\, \ep_{ijk} \, [e^0 \wedge e^i\wedge e^j
\wedge e^{\td k} + e^{\td 0}\wedge e^i \wedge e^{\td j} \wedge
e^{\td k}]\,,  \label{harm4} \eea
%%%%%
implying that $|L_{(2,2)}|^2 = \fft{360}{\cosh^8 r}\,.$

  The 8-dimensional Stenzel manifold can be used as the transverse
space for constructing a deformed M2-brane, by employing the
explicit construction in Subsection 3.1 and the explicit form
(\ref{harm4}).  The solution of (\ref{m2sol}) and (\ref{m2har}) can be
obtained by
making a coordinate redefinition, $2+ \cosh 2r = y^4$, in terms of which
$H$ is given by
%%%%%
\be 
H= c_0- \fft{5m^2\,(5y^5 - 7y)}{4\sqrt2 (y^4-1)^{3/2}} +
\fft{25m^2}{4\sqrt2} F(\arcsin(\fft1{y})|-1)\,, 
\end{equation} 
%%%%%%
where $c_0$ is an integration constant, and  $F(\phi|m) \equiv
\int_0^\phi (1- m\, \sin^2\theta)^{-1/2}\, d\theta\, $ is the
incomplete elliptic integral of the first kind.

   The function $H$ is regular for $r$ running 
from 0 to infinity. Near $r=0$, $H$ approaches 
a constant, and  at large $r$ we have
%%%%%
\be H = c_0 + \fft{640m^2}{2187\rho^6} - \fft{20480\, 2^{1/3}\,
m^2}{28431\, 3^{2/3}\, \rho^{26/3}} + \cdots\,, \end{equation} 
%%%%%
where $\rho$ is the proper distance, defined by $h\, dr =
d\rho$.  Thus the M2-brane has no singularity, and it has a
well-defined ADM mass.

\subsubsection{Properties of Resolved M2-branes}

  {\it (i) Supersymmetry} For the supersymmetry of the deformed
  solution to be preserved, the harmonic 4-form must satisfy 
\cite{bebe,hata}:
%%%%%
\be 
F_{abcd}\, \Gamma_{bcd}\, \eta=0\,.\label{d8susycon} \end{equation} 
%%%%%
where $\eta$ is a covariantly constant spinor in the Ricci flat
8-space. For $L{(2,2)}$ in the Stenzel metric, one can show  
that this is indeed satisfied for each such spinor.
In other words, turning on the deforming flux from the
harmonic 4-form $G_\4$ does not lead to any further breaking of
supersymmetry, and so the resolved M2-brane preserves $1/4$ of
the original supersymmetry.

   For other examples of the K\"ahler, hyper-K\"ahler, and old
$Spin(7)$ holonomy transverse spaces, $L^2$-normalizable 4-forms were
also constructed, that lead to supersymmetric M2-branes with the same
numbers of preserved supersymmetries as the M2-branes without the
additional field strengths.

 {\it (ii) Flux Integrals}
Interestingly, there are no additional conserved charges\footnote{
The issue of brane charges,
and their relation to the notion of brane
wrapping, is a subtle one, which we shall not pursue in detail here.
Thus, we shall just follow some standard terminology for now.}
for the resolved M2-brane. In particular, the resolved  M2-brane with
the transverse Stenzel metric has  $\int_{S^4} F_4\sim m
L_{(2,2}) \sim (\cosh r)^{-1}\,
\int_{S^4}\nu\wedge\sigma_1\wedge\sigma_2\wedge\sigma_3\to 0$
thus leading to zero  ``fractional'' charges. Here $S^4$ is
the bolt in the 8-dimensional Stenzel metric, and  had the above
integral been non-zero this would have implied non-zero
fractional charge for M5-branes wrapped around a  three-cycle
dual to $S^4$ within $M_7$.
The electic flux integral of the original M2-brane is determined
by $\int _{M_7} *F_{(4)}\sim m^2$, leading to a non-zero
 charge proportional to $m^2$. Here $M_7$ is
  the transverse seven-dimensional
base over which the resolved cone (of the Ricci flat space) is defined.

    All the known examples on resolved M2-branes
with transverse spaces that are asymptotically conical
turn out to have {\it zero fractional charge}.

{\it (iii) Dual Field Theory} The $L^2$-normalizability of the 4-forms
supporting additional fields-strengths of the resolved M2-branes imply
that in the decoupling limit the large distance space-time is still of
the form $AdS_4\times M_7$.  This indicates that in the ultraviolet
(UV) we shall get $D=3$ CFT, with an ${\cal N}=2$ field theory (${\cal
N}=1$ for the original $Spin(7)$ space and an ${\cal N}=3$ theory for the
hyper-K\"ahler space).  On the other hand, the additional field
strengths imply \cite{hekl} a deformation by relevant operators
associated with the pseudo-scalar fields in a dual field theory. 
This property is conjectured in \cite{hekl} to extend to all
resolved M2-branes whose Ricci flat transverse spaces are
asymptotically conical.

\subsection{Properties of fractional D2-branes}
Fractional D2-branes involve $G_2$ holonomy metrics, and
the examples are summarized for each of the original 
$G_2$ holonomy metrics in Subsection 4.2. In the following
subsections we summarize the properties of fractional branes for these
examples.\footnote{In the context of M-theory on $G_2$ holonomy
spaces the properties of harmonic three forms play an important
role as  axion moduli in $D=4$ ${\cal N}=1 $ field theory. For
details see \cite{atwi}.}

\subsubsection{ Cones over $S^2\times S^4$  and $S^2\times \CP^2$}

We can substitute  the explicit solution for $L_{(3)}$
(\ref{vform}) into (\ref{d2sol}) and obtain the solution for the
resolved fractional D2-brane.  It is difficult to give a fully
explicit expression for $H$, since the harmonic 3-form itself has
a rather complicated expression.  However,  the large and small
distance analysis of the solution is straightforward. 	In these limits
we have
%%%%%
\bea 
\hbox{Large distance}: && H = c_0 + \fft{Q}{r^5} -
\fft{m^2}{4 r^6}
 +\cdots \,,\nn\\
\hbox{Small distance}: && H=c_1 - \fft{11m^2\,
(r-1)}{24}+\cdots\,, 
\eea
%%%%%
where
%%%%%
\be 
Q= \fft{m^2}{30}\, \int_1^\infty dr\, r^4\, \sqrt{r^4-1}\,
|L_{(3)}|^2 \,. 
\end{equation} 
%%%%%
The solution is asymptotically of the same form as the D2-brane
without fractional charges, except that the D2-brane charge
$Q$ is now proportional to $m^2$. 
At short distance, on the other hand, the solution is
regular. Examining the flux associated with additional field
strength $F_{(4)}$  yields {\it a finite integral } 
$L_{(4)}\equiv {* L_{(3)}}= \ft14 u_1\, r^4\, \Omega_\4 +
\cdots\,,$ (where $\Omega_\4$ is the volume form of the unit
4-sphere and $u_1$ is given by (\ref{usol}))  over the 4-sphere at infinity:
%%%%%
\be 
P_4\equiv  \fft{1}{V(S^4)}\, \int L_{(4)} = \ft14\,. 
\end{equation} 
%%%%%
This  result implies that our D2-brane solution carries additional
wrapped D4-brane charge.

  This is a completely regular supersymmetric solution describing
the usual D2-brane together with a fractional D2-brane coming from
the wrapping of D4-branes around a 2-cycle. (There is an analogous
fractional D2-brane in which the Ricci-flat 7-manifold is
replaced by the related example where the principal orbits are
$S^2$ bundles over $\CP^2$ instead of $S^4$.)

   To summarise, because of the existence of a normalizable harmonic
3-form on these Ricci-flat 7-manifolds with the topology of
$\R^3$ bundles over $S^4$ or $\CP^2$, the corresponding
fractional D2-brane solutions are regular both at small distance
as well as at large distance.  Thus, in the decoupling limit they
have the same asymptotic large-distance behaviour as regular
D2-branes with Euclidean transverse spaces. As a consequence, the
dual asymptotic field theory is that of a regular D2-brane (whose
original charge $Q\sim N$ determines the SYM factor $SU(N)$), but
now with a different overall charge $Q\propto m^2\sim M$, which
is related to the contribution of the additional fluxes of the
wrapped D4-branes, and thus indicating a single SYM factor
$SU(M)$.  These deformed solutions preserve $1/16$ of the
original supersymmetry and so they describe a dual three-dimensional
${\cal N}=1$ field theory.

\subsubsection {Resolved cone over $S^3\times S^3$}
The solution
for the function $H$ in the corresponding deformed D2-brane
solution (\ref{d2sol}) was shown to be given by \cite{clp1}:
%%%%%%
\bea 
H&=& c_0 + \fft{m^2 (r+1) (16r^7 + 24 r^6 + 48  r^5 + 47
r^4\ + 54  r^3 + 36 r^2 + 18  r + 9)}{108 r^3\,
(r^2 + r + 1)^3}\,\nn\\
&&+ \fft{8 m^2}{27 \sqrt3}\, {\rm arctan} \Big[\fft{2r
+1}{\sqrt3}\Big] \,.  
\eea
%%%%%%%
The function $H$ is non-singular for $r$ running from the origin
at $r=1$ to infinity. (Note also that the small distance behaviour
of the above solution is qualitatively the same as that of the
fractional D2-brane discussed in the previous subsection, thus
indicating the same universal infrared behaviour of the dual field
theories for both types of solution.)

   The fractional flux is supported by an harmonic 3-form that
is not $L^2$ normalizable, and the solution, while regular at small
distances, corresponds at large distances to the resolution of the
D2-brane whose ``charge''-$Q$ is now given in terms of the parameter
$m$ by $Q= -4m^2/15 r$.  The deformed D2-brane solution also has a
non-vanishing flux for the NS-NS 5-brane, which wraps around the
$S^3$.  Specifically, at large $r$ we find that $L_{(3)} =
\ft1{3\sqrt3}\, v_3 \, \Omega_\3 + \cdots\,, $
%%%%%
(where $\Omega_3$ is the volume form and $v_3$ is defined in
(\ref{vsol})) and
this leads to a non-vanishing, finite $F_{(3)}$ flux, integrated
over the 3-sphere associated with the metric $\Sigma_i^2$ at
infinity,
%%%%%
\be 
P_3 \equiv  \fft1{V(S^3)}\, \int L_{(3)} = \fft1{\sqrt3}\,.
\end{equation} 
%%%%%
 This implies that the D2-brane carries additional
wrapped NS-NS 5-brane charge and  the solution can be viewed as a
fractional NS-NS 2-brane, together with the usual D2-brane
supported by the 4-form.

 The linear growth of the ``charge'' with distance accounts for the
asymptotic renormalization group running of $g_1^{-2}-g_2^{-2}$,
linearly with the energy scale, where $g_1$ and $g_2$ are the gauge
couplings of two SYM factors. The solution preserves $1/16$ of the
original supersymmetry, thus describing a regular supergravity dual of
a three-dimensional ${\cal N}=1$ field theory.

\subsection{Fractional M2-branes with new $Spin(7)$ holonomy
space}

One of the main motivations for constructing new $Spin(7)$ holonomy
metrics was to try to relate them to 7-metrics of $G_2$
holonomy, and thus in turn to find interesting relations between
fractional M2-branes and D2-branes.  Indeed
 the $Spin(7)$
holonomy metrics,  described in  Subsection 4.3, asymptotically
approach $S^1\times M_7$ locally, where the size of the circle is
{\it finite} and $M_7$ is asymptotically the $G_2$ holonomy
space, whose principal orbits are those of the $S^2$ bundle over
$S^4$.

The normalizable, supersymmetric 4-form implies a resolved
M2-branes solution. For the harmonic form, the function $H$ is given
by
%%%%%
\be H=1 + \fft{m^2(1323r^6 + 9786r^5 + 32937r^4 + 64428r^3 +
52237r^2 - 136934r + 29983)}{1680(r+1)^9(r-1)^2}\,.\label{h2}
\end{equation} 
%%%%%%%%
(Again for the sake of simplicity we set $\ell=1$.) The solution is
regular everywhere.

 For the manifold
obtained by the analytic continuation $\ell\to -\ell$, the solution is 
given by
%%%%%
\bea 
H= 1 + \fft{m^2(3r^2 +26r +63)}{20(r+1)^2(r+3)^5}\,.
\label{h1} 
\eea
%%%%%
The solution is smooth everywhere, and  it interpolates between
eleven-dimensional Minkowski spacetime at small distance and
$M_3\times S^1\times {\cal M}_7$ locally at large distance.

Note that this brane does have a fractional charge: the
4-form $F_\4$ carries a magnetic M5-brane charge proportional to $m$ in
addition to the electric M2-brane charge proportional $m^2$.  The
magnetic charge is given by $ Q_m =\fft{1}{\omega_4}\int F_\4 =
q\, m\,, $ where $\omega_4$ is the volume of the unit 4-sphere,
and $q=\ft12$  for the two cases discussed above. Thus, in
contrast to the resolved M2-branes with transverse
asymptotically flat spaces,  these resolved M2-brane solutions,
describe fractional magnetic M2-branes as wrapped M5-branes,
together with the usual electric M2-brane. This is because
the new $Spin(7)$ holonomy spaces are not
asymptotically conical, but instead have an asymptotically local
conical  structure with an $S^1$ whose radius tends to a constant
at infinity.

     The fractional D2-brane obtained from the wrapping of a D4-brane
around the $S^2$ in a manifold of $G_2$ holonomy was conjectured in
\cite{cglp3} to be related to a resolved M2-brane with a transverse
8-dimensional space of $Spin(7)$ holonomy. The examples found in
\cite{cglpspin7} and summarized above provide a concrete realization
of this conjecture: reducing the solution on the $U(1)$ fiber in
$Spin(7)$ holonomy space yields a regular D2-brane charge proportional
to $m^2$ (coming from the double dimensional reduction of the
M2-brane), and a fractional D2-brane charge proportional to $m$
arising from D4-branes wrapped on 2-cycles (coming from the vertical
reduction of the M5-brane), as well as a wrapped D6-brane charge
proportional to $\ell$. The D6-brane charge is a consequence of the
Kaluza-Klein reduction on the $U(1)$ fiber in the original Ricci-flat
space, corresponding to a reduction from $D=11$ M-theory to $D=10$
Type IIA supergravity \cite{ac,atmava,ednu}.

\section{Conclusions  and open avenues}

In this contribution we  have  presented a summary of some recent
developments in the construction of regular $p$-brane configurations with
less than maximal supersymmetry. In particular, the method
involves the  introduction of complete non-compact special holonomy
metrics and additional fluxes, supported by harmonic-forms in 
special holonomy spaces,   which modify the original $p$-brane
solutions via Chern-Simons ({\it transgression}) terms.

   The work led to a number of important {\it mathematical
developments}  which we have also summarized.  Firstly, the
construction of harmonic forms for special holonomy spaces in
diverse dimensions
 was reviewed, and the  explicit construction of harmonic forms
for Stenzel metrics was summarized. In particular, only middle
dimension  harmonic forms of the type $(n,n)$ in $D=4n$ dimensions
are $L^2$ normalizable, while  those of type $(n+1,n)$ in
$D=4n+2$ are marginally non-$L^2$ normalizable.  Secondly, a
construction of new two-parameter $Spin(7)$ holonomy spaces was
discussed.  These have the
property that they interpolate asymptotically between a local $S^1\times
{\cal M}_7$, where the length of the circle is finite and ${\cal
M}_7$ is the $G_2$ holonomy space with the topology of the $S^2$
bundle over $S^4$, while at small distance  they
approach the ``old'' $Spin(7)$ holonomy space with the topology of
the chiral spin bundle  over $S^4$.

   The mathematical developments also led to a number of important
{\it physics implications}, relevant for the properties of the
resolved $p$-brane solutions. In particular, in this contribution we
focused on the properties of resolved M2-branes with 8-dimensional
special holonomy transverse spaces (e.g., Stenzel, hyper-K\"ahler and
$Spin(7)$ holonomy spaces) and the fractional D2-branes with three
7-dimensional $G_2$ holonomy transverse spaces.  Resolved M2-branes
are always supported by $L^2$-normalizable harmonic forms and thus
they are regular at short distance and have decoupling limits at large
distance that yield $AdS_4$.  They have no conserved additional
(fractional) charges, but only the usual M2-brane electric charge
(which is now proportional to $m^2$, where $m$ is the strength of the
additional flux). The dual 3-dimensional dual field theory is therefore
a superconformal field theory (with ${\cal N}=1$ or ${\cal
N}=2$ supersymmetry) which is in turn perturbed by marginal operators
associated with pseudo-scalar fields \cite{hekl}. On the other hand
the fractional D2-branes are either supported by $L^2$ normalizable
harmonic forms (in the case of $G_2$ holonomy spaces which are $\R^3$
bundles over $S^4$ or $\CP^2$) or by a harmonic form that has
a linearly divergent integrated norm (in the case of the $\R^4$ bundle
over $S^3$). In all these cases there are
conserved fractional charges corresponding to D4-branes wrapping the
2-cycles dual to $S^4$ ($\CP^2$) in ${\cal M}_7$, or to NS-NS
5-branes wrapping the 3-cycle dual to $S^3$ in ${\cal M}_7$,
respectively.

   An interesting implication involves the properties of fractional
M2-branes using the new $Spin(7)$ holonomy
spaces.  After reduction on $S^1$ these give fractional D2-branes
where, in addition to the original D2-brane charge, there is fractional
magnetic charge for D4-branes wrapping 2-cycles {\it and} for 
D6-branes wrapping 4-cycles, arising
as a consequence of the Kaluza-Klein reduction. The fact that the
resolved M2-brane on the new spin $Spin(7)$ holonomy space has 
non-zero fractional charge is a consequence of the asymptotically
locally conical structure of the new
$Spin(7)$ holonomy space.

   There are a number of  {\it open avenues} in the exploration of
further properties of such solutions. In particular, it is of
importance to study the properties of the dual three dimensional field
theories in greater detail.

%   Another important direction involves the study of novel special
% holonomy spaces in M-theory. In particular, new constructions of $G_2$
% holonomy spaces \cite{bggg,cglp5} are of importance in the study of
% ${\cal N}=1$ $D=4$ field theory aspects of M-theory
% \cite{atwi,ac,atmava}. Recently, the study of M-theory on spaces of
% $G_2$ holonomy has attracted considerable attention. In particular, it
% has been proposed that M-theory compactified on a certain singular
% seven-dimensional space with $G_2$ holonomy might be related to an
% ${\cal N}=1$, $D=4$ gauge theory \cite{wi,atwi,ac,atmava,agva} that
% has no conformal symmetry.  The quantum aspects of M-theory dynamics
% on spaces of $G_2$ holonomy can provide insights into non-perturbative
% aspects of four-dimensional ${\cal N}=1$ field theories, such as the
% preservation of global symmetries, and phase transitions.  For example,
% Ref. \cite{atwi} provides an elegant exposition and study of these
% phenomena for the three manifolds of $G_2$ holonomy that were obtained
% in \cite{brysal,gibpagpop}.
% 
%     A new exciting development in this direction is the discovery of
% M3-brane configurations \cite{clpmassless,cglp5} that have a flat
% 4-dimensional world-volumes, and transverse spaces that are
% deformations of $G_2$ manifolds, with non-vanishing 4-form field
% contributions.  An investigation of their properties is a subject for
% further research.  These configurations turn out to have zero charge
% and ADM mass (leading to naked singularities at small distances).
% 
Another important direction involves the study  of novel special
holonomy  spaces in  M-theory. In particular, novel constructions
of $G_2$ holonomy spaces \cite{cglp5,bggg} are of importance in
the study of ${\cal N}=1$ D=4  field theory aspects of  M-theory
\cite{atwi,ac,atmava} and the study of M-theory on spaces of
$G_2$ holonomy has recently attracted considerable attention.
Specifically,  it has been proposed that M-theory compactified on a
certain singular seven-dimensional space with $G_2$ holonomy
might be related to an ${\cal N}=1$, $D=4$ gauge theory
\cite{wi,atwi,ac,atmava,agva} that has no conformal symmetry.
 The
quantum aspects of M-theory dynamics on spaces of $G_2$ holonomy
can provide insights into non-perturbative aspects of
four-dimensional ${\cal N}=1$ field theories, such as the
preservation of global symmetries and phase transitions.  For
example, Ref. \cite{atwi} provides an elegant exposition and study
of these phenomena for the three manifolds of $G_2$ holonomy that
were obtained in \cite{brysal,gibpagpop}.

 A new exciting development  in this
direction  is  the discovery of M3-brane configurations
\cite{clpmassless,cglp5} which have
a flat 4-dimensional world-volume  and the transverse space that
is a deformation of the $G_2$ along with the 4-form
field strength  turned on.
 These configurations turn
out to have zero charge and ADM mass (leading to naked
singularities at small distances). A study of their properties is the
subject of  further research.

\section*{Acknowledgement}

 M.C. thanks the organizers of the M-theory workshop at the Institute
of Theoretical Physics at the University of California in Santa
Barbara and Center for Applied Mathematics and Theoretical Physics in
Maribor, Slovenia, and C.N.P. thanks DAMTP and St. John's College
Cambridge, for hospitality during the completion of this contribution.
Research is supported in part by DOE grant DE-FG02-95ER40893, NSF
grant No. PHY99-07949, Class of 1965 Endowed Term Chair and NATO
Collaborative reserach grant 976951 (M.C.), in full by DOE grant
DE-FG02-95ER40899 (H.L.) and in part by DOE grant DE-FG03-95ER40917
(C.P.).


\begin{thebibliography}{99}
\bm{klst} I.R. Klebanov and M.J. Strassler, {\sl Supergravity and a
confining gauge theory: duality cascades and $\chi$SB-resolution of
naked singularities}, JHEP {\bf 0008}, 052 (2000), [hep-th/0007191].


\bm{clp1} M. Cveti\v{c}, H. L\"u and C.N. Pope, {\it Brane resolution
through transgression}, Nucl.\ Phys.\ B {\bf 600},
103 (2001) [hep-th/0011023].

\bm{cglp1} M. Cveti\v{c}, G.W. Gibbons, H. L\"u and C.N. Pope, {\it
Ricci-flat metrics, harmonic forms and brane resolutions}, hep-th/0012011.

\bm{cglp3} M.~Cveti\v c, G.~W.~Gibbons, H.~L\"u and C.~N.~Pope,
{\it Hyper-K\"ahler Calabi metric, $L^2$ harmonic forms, resolved
M2-branes, and AdS$_4$/CFT$_3$ correspondence}, hep-th/0102185.

\bm{cglpspin7}
M.~Cveti\v c, G.~W.~Gibbons, H.~L\"u and C.~N.~Pope,
{\it New complete non-compact Spin(7) manifolds},
hep-th/0103155.
%%CITATION = HEP-TH 0103155;%%

\bm{cglpspin7m} M. Cveti\v c, G.W. Gibbons, H. L\"u and C.N. Pope,
{\it New cohomogeneity one metrics with Spin(7)
 holonomy}, math.DG/0105119.

\bm{cglp2}
M.~Cveti\v c, G.~W.~Gibbons, H.~L\"u and C.~N.~Pope,
{\it Supersymmetric non-singular fractional D2-branes and NS-NS 2-branes},
hep-th/0101096, to appear in Nucl. Phys. {\bf B}.


\bibitem{ma} J. Maldacena, {\sl The large $N$ limit of
superconformal field theories and supergravity},
Adv. Theor. Math. Phys. {\bf 2} (1998) 231, hep-th/9711200.

\bibitem{gkp} S.S. Gubser, I.R. Klebanov and A.M. Polyakov, {\sl Gauge
theory correlators from non-critical string theory}, Phys. Lett. {\bf
B428} (1998) 105, hep-th/9802109.

\bibitem{wi} E. Witten, {\sl Anti-de Sitter space and holography},
Adv. Theor. Math. Phys. {\bf 2} (1998) 253, hep-th/980215.

\bibitem{becv}
K.~Behrndt and M.~Cveti\v c,
{\it Gauging of N = 2 supergravity hypermultiplet and novel
renormalization group flows,}
hep-th/0101007.
 
\bibitem{ceetal}
A.~Ceresole, G.~Dall'Agata, R.~Kallosh and A.~Van Proeyen,
{\it Hypermultiplets, domain walls and supersymmetric attractors,}
hep-th/0104056.

\bm{grpo} M. Gra\~na and J. Polchinski, {\it Supersymmetric
three-form flux perturbations on AdS$_5$}, hep-th/0009211.

\bm{manu} J.~Maldacena and C.~Nu\~nez, {\sl Supergravity
description of field theories on curved manifolds and a no  go
theorem,} Int.\ J.\ Mod.\ Phys.\ A {\bf 16}, 822 (2001)
[hep-th/0007018].

\bm{gu} S.S. Gubser, {\it Supersymmetry and F-theory
realization of the deformed conifold with three-form flux},
hep-th/0010010.


\bm{pats} L.A. Pando Zayas and A.A. Tseytlin, {\it 3-branes on a
resolved conifold}, hep-th/0010088.

\bm{bebe} K. Becker and M. Becker, {\sl Compactifying M-theory to
four dimensions,} JHEP {\bf 0011}, 029 (2000) [hep-th/0010282].

\bm{bvflmp} M. Bertolini, P. Di Vecchia, M. Frau, A. Lerda, R.
Marotta and I. Pesando, {\sl Fractional D-branes and their gauge
duals}, hep-th/0011077.

\bm{ah} O. Aharony, {\it A note on holgraphic interpretation of
string theory backgrounds  with varying  flux}, hep-th/0101013.



\bm{ca} E.~Caceres and R.~Hernandez,
{\it Glueball masses for the deformed conifold theory},
Phys.\ Lett.\ B {\bf 504}, 64 (2001)
[hep-th/0011204].


\bm{gaetal} J.~P.~Gauntlett, N.~Kim and D.~Waldram, {\sl
M-fivebranes wrapped on supersymmetric cycles,} hep-th/0012195.

\bm{klwi1} I.R. Klebanov and E. Witten, {\it Superconformal field
theory on threebranes at a Calabi-Yau singularity}, Nucl. Phys.
{\bf B536}, 199 (1998) [hep-th/9807080].

%\bm{klwi2} I. Klebanov and E. Witten,




\bm{gukl} S.S. Gubser and I.R. Klebanov, {\sl Baryons and domain
walls in an $N=1$ superconformal gauge theory}, Phys. Rev. {\bf
D58}, 125025 (1998), [hep-th/9808075].


\bm{klne} I.R. Klebanov and N. Nekrasov, {\sl Gravity duals of
fractional branes and logarithmic RG flow}, Nucl. Phys. {\bf B574},
263 (2000), [hep-th/9911096].

\bm{klts} I.R. Klebanov and A.A. Tseytlin, {\sl Gravity duals of
supersymmetric $SU(N)\times SU(N+m)$ gauge theories}, Nucl. Phys.
{\bf B578}, 123 (2000) [hep-th/0002159].

\bm{egha} T. Eguchi and A.J. Hanson, {\it Asymptotically flat
self-dual solutions to Euclidean gravity}, Phys. Lett. {\bf B74},
249 (1978).

\bm{bebe0}K.~Becker and M.~Becker,
{\it M-theory on eight-manifolds}, 
Nucl.\ Phys.\ B {\bf 477}, 155 (1996)
[hep-th/9605053].


\bm{duetal} M.J. Duff, J.M. Evans, R.R. Khuri, J.X. Lu and R. Minasian,
{\sl The octonionic membrane,} Phys.\ Lett.\  {\bf B412},
281 (1997) [hep-th/9706124].

\bm{hata} S.W. Hawking and M.M. Taylor-Robinson, {\it Bulk charges in
eleven dimensions}, Phys. Rev. {\bf D58} 025006
(1998) [hep-th/9711042].

% \bm{gss} B.R. Greene, K. Schalm and G. Shiu, {\sl Warped
% compactifications in M and F theory,} Nucl.\ Phys.\ {\bf B584},
% 480 (2000) [hep-th/0004103].



\bm{kbe} K. Becker, {\it A note on compactifications on
Spin(7)-holonomy manifolds},\\
hep-th/0011114.

\bm{hekl} C.P. Herzog and I.R. Klebanov, {\sl Gravity duals of
fractional branes in various dimensions}, hep-th/0101020.

\bm{he}C.~P.~Herzog and P.~Ouyang,
{\it Fractional D1-branes at finite temperature},
hep-th/0104069.

\bm{dasw} A.S. Dancer and  A. Swann, {\it Hyperk\"ahler metrics
of cohomogeneity one},  J. Geometry and Physics, {\bf 21}, 218
(1997).

\bm{brysal} R.L. Bryant and S. Salamon, {\sl On the construction
of some complete metrics with exceptional holonomy}, Duke Math.
J. {\bf 58}, 829 (1989).


\bm{gibpagpop} G.W. Gibbons, D.N. Page and C.N. Pope, {\sl
Einstein metrics on $S^3$, $\R^3$ and $\R^4$ bundles}, Commun.
Math. Phys. {\bf 127}, 529 (1990).

\bm{cglp5} M. Cveti\v c, G.W. Gibbons, H. L\"u and C.N. Pope,
{\it Supersymmetric M3-branes and $G_2$ manifolds}, hep-th/0106031.

\bm{bggg} A. Brandhuber, J. Gomis, S.S. Gubser and S. Gukov, {\it
Gauge Theory at Large  N and New $G_2$ Holonomy Metrics},
hep-th/0106034.

\bm{st} M.B. Stenzel, {\it Ricci-flat metrics on the
complexification of a compact rank one symmetric space},
Manuscripta Mathematica, {\bf 80}, 151 (1993).



\bm{dloc} P.~Candelas and X.~C.~de la Ossa, {\it Comments On
Conifolds}, Nucl.\ Phys.\ B {\bf 342}, 246 (1990).
%%CITATION = NUPHA,B342,246;%%d


\bm{atwi} M.F. Atiyah and E. Witten, {\it M-theory Dynamics on a
Manifold of $G_2$ Holonomy}, to appear.


\bm{ac} B.~S.~Acharya, {\sl On realising $N = 1$ super Yang-Mills
in M theory,} hep-th/0011089.


\bm{atmava} M.~Atiyah, J.~Maldacena and C.~Vafa, {\sl An M-theory
flop as a large $n$ duality,} hep-th/0011256.

\bm{ednu} J.~D.~Edelstein and C.~Nu\~nez, {\it D6 branes and
M-theory geometrical transitions from gauged  supergravity}, JHEP
{\bf 0104}, 028 (2001) [hep-th/0103167].

\bm{agva} M.~Aganagic and C.~Vafa, {\it Mirror symmetry and a G(2)
flop},  hep-th/0105225.

\bm{clpmassless}  M. Cveti\v{c}, H. L\"u and C.N. Pope, {\it Massless
3-branes in M-theory}, hep-th/0105096.




\end{thebibliography}
\end{document}